\documentclass{osa-article}

\newcommand\eps{\varepsilon}
\newcommand\ee{\mathrm{e}}
\newcommand\ii{\mathrm{i}}
\newcommand\bmat[1]{\begin{bmatrix}#1\end{bmatrix}}
\newcommand\lt\left
\newcommand\rt\right
\newcommand\mat[1]{\mathbf{#1}}

\renewcommand{\Re}{\mathop{\mathrm{Re}}\nolimits}
\renewcommand{\Im}{\mathop{\mathrm{Im}}\nolimits}

\journal{osajournal}

\setprjcopyright

\articletype{Research Article}

\begin{document}

\title{Bound states in the continuum and high-Q resonances supported by a dielectric ridge on a slab waveguide}

\author{Evgeni~A.~Bezus,\authormark{1,2,*} Dmitry~A.~Bykov,\authormark{1,2} and Leonid~L.~Doskolovich\authormark{1,2}}

\address{\authormark{1}Image Processing Systems Institute -- Branch of the Federal Scientific Research Centre ``Crystallography and Photonics'' of Russian Academy of Sciences, 151 Molodogvardeyskaya st., Samara 443001, Russia\\}
\address{\authormark{2}Samara National Research University, 34 Moskovskoe shosse, Samara 443086, Russia\\}

\email{\authormark{*}evgeni.bezus@gmail.com}


\begin{abstract}
We investigate the diffraction of guided modes of a dielectric slab waveguide on a simple integrated structure consisting of a single dielectric ridge on the surface of the waveguide.
Numerical simulations based on aperiodic rigorous coupled-wave analysis demonstrate the existence of sharp resonant features and bound states in the continuum (BICs) in the reflectance and the transmittance spectra occurring at oblique incidence of a TE-polarized guided mode on the ridge.
Using the effective index method, we explain the resonances by the excitation of the cross-polarized modes of the ridge.
The formation of the BICs is confirmed using a theoretical model based on the coupled-wave theory.
The model suggests that the BICs occur due to coupling of quasi-TE and quasi-TM modes of the structure.
Simple analytical expressions for the angle of incidence and the ridge width predicting the location of the BICs are obtained.
The existence of high-Q resonances and BICs makes the considered integrated structure promising for filtering, sensing, transformation of optical signals, and enhancing nonlinear light--matter interactions.
\end{abstract}

\section{Introduction}\label{sec:intro}

In a wide class of planar (integrated) optoelectronic systems, processing of the optical signal is performed in a slab waveguide~\cite{haus,mossberg,dph1,dph2,dph3,dph4,echelle,tgpd,tgpf,hammer1,hammer2}.
Such geometry corresponds to the ``insulator-on-insulator'' platform and is suitable for the creation of fully integrated optical devices.
In this case, the processed signal corresponds to a superposition of slab waveguide modes having different propagation directions (in the case of spatial processing) or different frequencies (in the case of spectral processing).
In this regard, the design of integrated resonant photonic structures is of great interest~\cite{Miroshnichenko, Zhou}.

In the last few years, much attention has been paid to the investigation of the so-called bound states in the continuum (BICs) in resonant photonic structures (see the recent review~\cite{HsuReview} and references therein).
The BICs are the eigenmodes that have an infinite lifetime (and an infinitely high quality factor), although supported by a structure having open scattering channels.
The mode leakage to these channels is canceled either due to symmetry reasons or by means of parameter tuning~\cite{HsuReview}.
In photonics, BICs were studied in periodic structures (diffraction gratings, photonic crystal slabs, and arrays of dielectric rods or spheres)~\cite{Marinica, Monticone, Sadrieva, Sadrieva2, Hsu, Sadreev2, Blanchard},
defects and interfaces of photonic crystals~\cite{Hsu2, Sadreev1, Sadreev3},
and leaky optical waveguides~\cite{Zou}, among others~\cite{HsuReview}.
The structures supporting BICs can be used as ``unconventional'' narrow band waveguides \cite{Zou}.
A slight deviation from the BIC condition enables obtaining very high-Q resonators, which leads to various potential applications including lasers, sensors, and filters.

In this work, we investigate the resonant optical properties of a very simple structure consisting of a single subwavelength or near-wavelength ridge on the surface of a single-mode dielectric slab waveguide.
We analytically and numerically demonstrate that in the case of diffraction of a TE-polarized mode of the waveguide, the studied structure exhibits BICs and high-Q resonances associated with the excitation of the cross-polarized modes of the ridge.
Comparing to paper~\cite{Zou}, we present a theoretical model, which accurately describes the optical properties of the studied structure. In particular, we prove the existence of robust BICs and obtain simple closed-form expressions predicting their locations.
The discovered high-Q resonances and BICs supported by the presented planar structure make it promising for various applications including spatial (angular) and spectral filtering, transformation of optical signals, sensing, and enhancing nonlinear light--matter interactions.

The paper is organized as follows.
In Section~\ref{sec:2}, we define the geometry and discuss the scattering channels of the considered integrated structure.
In Section~\ref{sec:3}, we present full-wave numerical simulation results demonstrating the existence of high-Q resonances in the reflectance and transmittance spectra.
Using the effective index method, we qualitatively explain the formation of the resonances.
By rigorously calculating the quality factor of the eigenmodes of the structure, we show that it supports bound states in the continuum.
Section~\ref{sec:CWT} is dedicated to the derivation of a theoretical model describing the optical properties of the structure and explaining the BIC formation mechanism.
Section~\ref{sec:conclusion} concludes the paper.

\section{Geometry and scattering channels}\label{sec:2}

\begin{figure}[ht!]
\centering\includegraphics{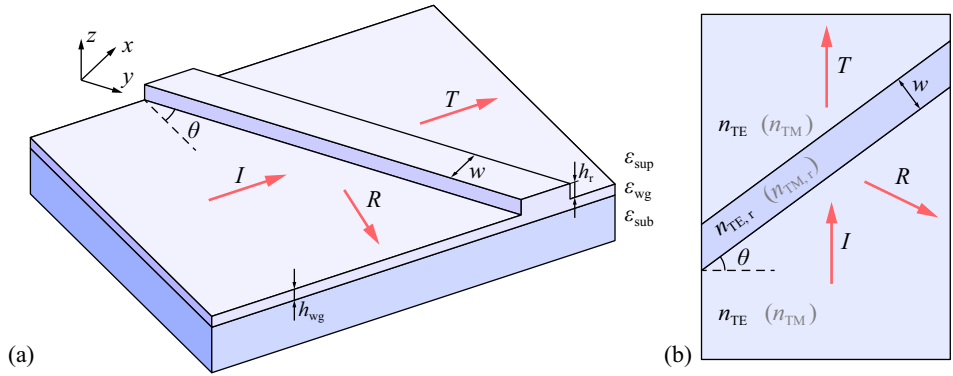}
\caption{\label{fig:1}(a)~Geometry of the considered integrated structure.
$I$, $R$, and $T$ denote the incident, reflected, and transmitted TE-polarized guided modes, respectively.
(b)~2D effective index model and the corresponding refractive indices.
}
\end{figure}

Let us consider a simple integrated structure shown in Fig.~\ref{fig:1}(a).
The structure consists of a dielectric ridge with the height $h_\mathrm{r} > 0$ and the width $w$ located on the surface of a single-mode dielectric slab waveguide with the thickness $h_\mathrm{wg}$.
For simplicity, we consider the case when the waveguide core and the ridge on its surface are made of the same material with dielectric permittivity $\eps_\mathrm{wg}$. 
Additionally, we assume that the substrate is optically denser than the superstrate: $\eps_\mathrm{sub} > \eps_\mathrm{sup}$.

We study the diffraction of a TE-polarized guided mode with an effective refractive index $n_\mathrm{TE}$ by an ``inclined'' ridge as shown in Fig.~\ref{fig:1}(a).
The angle between the ridge and the wave front of the incident mode is denoted by $\theta$ and referred to as the angle of incidence.

In the general case, reflected and transmitted TE- and TM-polarized guided modes are generated upon diffraction of the incident mode on the ridge, as well as a continuum of non-guided waves in the superstrate and substrate propagating away from the waveguide core layer.
However, it was recently demonstrated that, under certain conditions, the polarization conversion and out-of-plane scattering in such ``integrated'' diffraction problems can be completely eliminated~\cite{bezus_spp, hammer1, hammer2}.

The considered structure is invariant to translation in the direction parallel to the ridge interfaces (in the $y$-direction). According to Maxwell's equations, the corresponding tangential wave vector component of the incident wave $k_y = k_0 n_\mathrm{TE} \sin\theta$ has to be conserved for all the outgoing waves.
Here, $k_0 = 2\pi/\lambda$ is the wave number and $\lambda$ is the free-space wavelength.
Therefore, if the incidence angle is chosen so that $k_y^2 > k_0^2 \eps_\mathrm{sub}$, this tangential component has greater magnitude than that of the wave vector of the propagating plane waves in the substrate.
It means that at angles of incidence greater than the ``critical'' angle
\begin{equation}
\label{crsub}
 \theta_\mathrm{sub} = \arcsin\big(\sqrt{\eps_\mathrm{sub}}/n_\mathrm{TE}\big),
\end{equation}
all the waves scattered to the substrate are evanescent and do not carry energy. 
A similar expression can be written for the critical angle $\theta_\mathrm{sup}$ describing the scattering to the superstrate.
The assumed inequality $\eps_\mathrm{sub} > \eps_\mathrm{sup}$ implies that $\theta_\mathrm{sub} > \theta_\mathrm{sup}$, and therefore no out-of-plane scattering to the superstrate occurs once $\theta > \theta_\mathrm{sub}$.

A similar approach can be applied to eliminate TM-polarized reflected and transmitted modes (with effective refractive index $n_\mathrm{TM}$) generated upon diffraction of the incident TE-polarized mode.
Indeed, since in a single-mode slab waveguide $n_\mathrm{TE} > n_\mathrm{TM}$~\cite{lifante}, a cutoff angle for the TM-polarized modes exists.
Similarly to the derivation of Eq.~\eqref{crsub}, on can show that at angles of incidence greater than
\begin{equation}
\label{crTM}
\theta_\mathrm{TM} = \arcsin(n_\mathrm{TM}/n_\mathrm{TE}),
\end{equation}
no propagating outgoing TM-polarized modes are generated.

Note that $n_\mathrm{TM} > \sqrt{\eps_\mathrm{sub}}$, therefore, $\theta_\mathrm{TM} > \theta_\mathrm{sub}$. Hence, at $\theta > \theta_\mathrm{TM}$, only two scattering channels remain open in the considered structure, namely, the reflected and transmitted TE-polarized modes.

We obtained the cutoff conditions for the waveguide of thickness $h_\mathrm{wg}$ concerning the outgoing waves in the considered diffraction problem.
In a similar way, the ridge region can be considered as a segment of a slab waveguide with a greater thickness $h_\mathrm{wg} + h_\mathrm{r}$.
Denoting the effective index of the TM-polarized mode of this slab waveguide by $n_\mathrm{TM, r}$, we can write the corresponding cutoff angle as
\begin{equation}
\label{crTMr}
\theta_\mathrm{TM,r} = \arcsin(n_\mathrm{TM,r}/n_\mathrm{TE}).
\end{equation}
Therefore, in the angular range $\theta_\mathrm{TM} < \theta < \theta_\mathrm{TM,r}$ the ridge region supports both TE- and TM-polarized modes, while the scattered field contains only two TE-polarized modes. 
As we show below, it is in this region that the structure exhibits remarkable optical properties.

\section{Numerical simulations: high-Q resonances and BICs}\label{sec:3}

In this section, we numerically study the optical properties of the considered integrated structure.
The simulations were carried out for the following parameters: free-space wavelength $\lambda = 630$~nm, waveguide thickness $h_\mathrm{wg} = 80$~nm, ridge height $h_\mathrm{r} = 30$~nm, dielectric permittivities of the superstrate, waveguide layer, and substrate $\eps_\mathrm{sup} = 1$, $\eps_\mathrm{wg} = 3.3212$ (GaP), and $\eps_\mathrm{sub} = 1.45$, respectively.

By solving the dispersion equation of a slab waveguide~\cite{pocket}, let us first calculate the effective refractive indices and the cutoff angles [Eqs.~\eqref{crsub}--\eqref{crTMr}] discussed in the previous section.
At the considered parameters, the slab waveguide outside the ridge (thickness $h_\mathrm{wg}$) supports a TE-polarized mode with effective refractive index $n_\mathrm{TE} = 2.5913$ and a TM-polarized mode with effective refractive index $n_\mathrm{TM} = 1.6327$.
The slab waveguide corresponding to the ridge region (thickness $h_\mathrm{wg} + h_\mathrm{r}$) supports TE- and TM-polarized modes with effective refractive indices $n_\mathrm{TE,r} = 2.8192$ and $n_\mathrm{TM,r} = 2.1867$, respectively.
Critical angles of incidence corresponding to the closing of different scattering channels discussed above amount to 
$\theta_\mathrm{sup} = 22.7^\circ$ (out-of-plane scattering to the superstrate), 
$\theta_\mathrm{sub} = 34.0^\circ$ (out-of-plane scattering to the substrate), and 
$\theta_\mathrm{TM} = 39.1^\circ$ (scattering to the TM-polarized guided mode).
The cutoff angle of the TM-polarized mode in the ridge region equals $\theta_\mathrm{TM,r} = 57.6^\circ$.

\subsection{Transmittance and reflectance spectra}\label{ssec:tandr}
\begin{figure}[ht!]
\centering\includegraphics{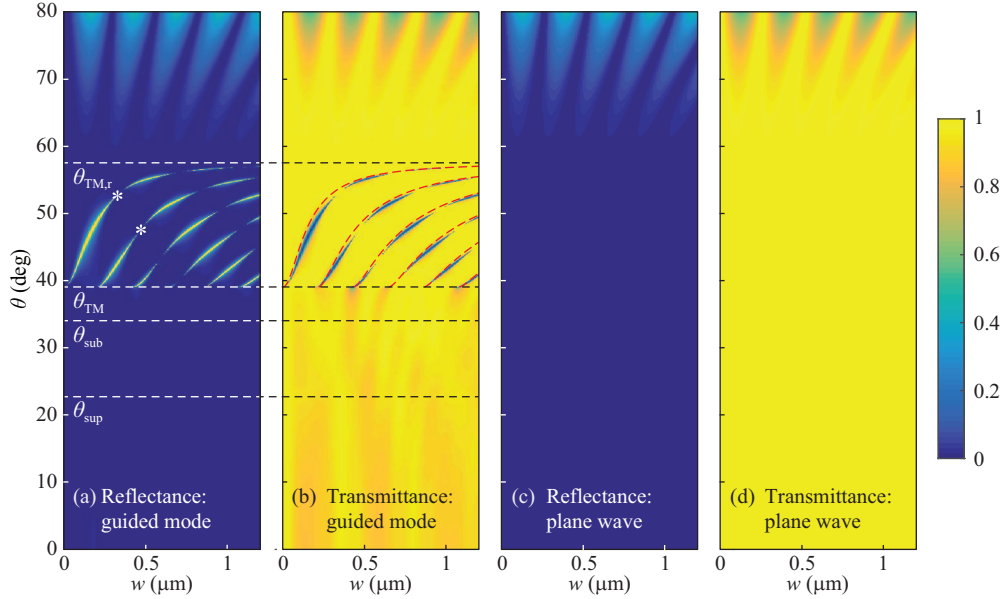}
\caption{\label{fig:2}Reflectance~(a) and transmittance~(b) of the TE-polarized mode vs.\ the angle of incidence $\theta$ and the ridge width $w$. 
Horizontal dashed lines show the cutoff angles $\theta_\mathrm{sup}$, $\theta_\mathrm{sub}$, $\theta_\mathrm{TM}$, and $\theta_\mathrm{TM,r}$.
The field distributions in the structure at the points marked with asterisks in~(a) are shown in Fig.~\ref{fig:3}.
Dashed red curves in~(b) show the dispersion of quasi-TM modes calculated used the approximate Eq.~\eqref{eq:disprel}.
Reflectance~(c) and transmittance~(d) of a plane wave upon refraction by a thin film (the effective index model) are also shown.
}
\end{figure}

Figures~\ref{fig:2}(a) and~\ref{fig:2}(b) show the reflectance and transmittance of a TE-polarized mode upon diffraction by a ridge on the waveguide surface vs. the angle of incidence~$\theta$ and the ridge width~$w$. 
Numerical simulations were performed using an efficient in-house implementation of the conical diffraction formulation of the rigorous coupled-wave analysis method (RCWA)~\cite{rcwa1,rcwa2} extended to integrated optics problems~\cite{rcwa3,pml}.
The RCWA, also called the Fourier modal method, is an established numerical technique for solving Maxwell's equations.

For comparison, Figs.~\ref{fig:2}(c) and~\ref{fig:2}(d) show the reflectance and transmittance of a plane wave upon refraction by a thin film with the thickness equal to the ridge width $w$.
In this plane-wave diffraction problem, effective refractive indices of the TE-polarized modes in the corresponding regions of the integrated structure are used as the refractive indices of the superstrate and substrate ($n_\mathrm{TE}$) and of the film ($n_\mathrm{TE,r}$) [see Fig.~\ref{fig:1}(b)].
The polarization of the incident plane wave coincides with the polarization of the incident mode in the ``integrated'' problem (i.e. the plane wave is polarized in the plane of incidence).
The model diffraction problem of Fig.~\ref{fig:1}(b) is very similar to the effective index method (EIM), which is widely used for quick estimation of the effective refractive indices of the modes of dielectric photonic wire and rib waveguides~\cite{pollock,pocket}.

It is evident from Fig.~\ref{fig:2} that at $\theta > \theta_\mathrm{TM,r}$, the reflectance and transmittance of the guided mode (obtained using the RCWA) are very close to the reflectance and transmittance of the plane wave (obtained within the EIM approach).
The periodically arranged reflectance and transmittance extrema visible in this region are due to Fabry-P\'erot resonances of the TE-polarized mode inside the ridge.
In the region $\theta < \theta_\mathrm{TM}$, some discrepancies in the transmittance are present, which are caused by scattering to the reflected and transmitted cross-polarized (TM-polarized) modes and to non-guided waves in the substrate (at $\theta < \theta_\mathrm{sub}$) and in the superstrate (at $\theta < \theta_\mathrm{sup}$).
Nevertheless, in the considered angular range ($\theta < \theta_\mathrm{TM}$ or $\theta > \theta_\mathrm{TM,r}$) 
the EIM can be used to obtain reasonable estimates of the solution of the ``integrated'' diffraction problem.

At angles of incidence $\theta_\mathrm{TM} < \theta < \theta_\mathrm{TM,r}$, the situation is drastically different.
In this angular range, sharp resonant peaks and dips, which are not explained by the plane-wave model, are present in the reflectance and transmittance spectra shown in Figs.~\ref{fig:2}(a) and~\ref{fig:2}(b), respectively.

\subsection{High-Q resonances}
Starting from this subsection, we focus on the angular range $\theta_\mathrm{TM} < \theta < \theta_\mathrm{TM,r}$, where the structure exhibits the most interesting optical properties, namely, narrow resonances with high quality factor. 

\begin{figure}[ht!]
\centering\includegraphics{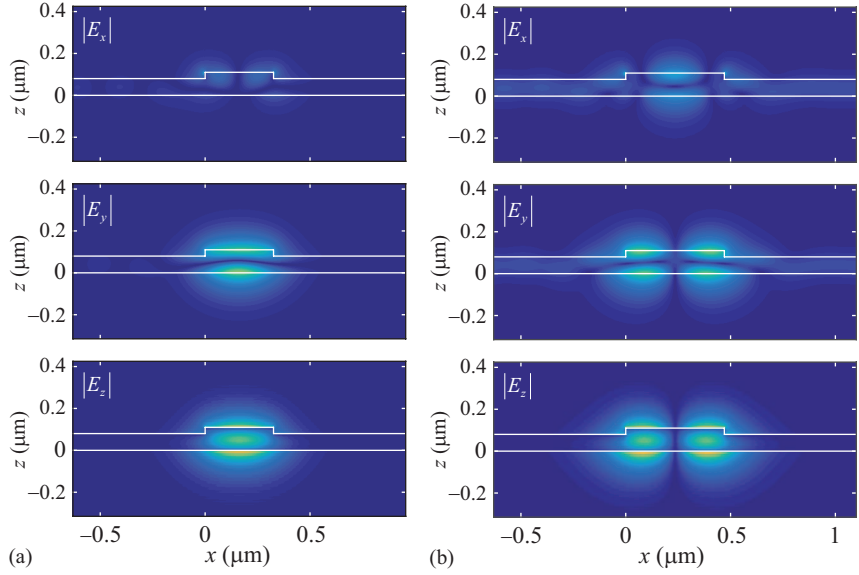}
\caption{\label{fig:3} Electric field distributions in the structure at resonance conditions:
(a)~$w = 326$ nm, $\theta = 52.66^\circ$; (b)~$w = 470$~nm, $\theta = 47.58^\circ$.}
\end{figure}

To reveal the nature of these high-Q resonances, let us investigate the field distribution in the structure at resonance conditions.
The electric field distributions in the integrated structure corresponding to two points depicted with white asterisks in Fig.~\ref{fig:2}(a) are shown in Figs.~\ref{fig:3}(a) and~\ref{fig:3}(b).
It is evident from Fig.~\ref{fig:3} that at resonances, the field distributions possess a strong $E_z$ component, which is absent in the incident TE-polarized mode.
This, along with the fact that the resonance region is bounded by the cutoff angles of the TM-polarized modes outside and inside the ridge, suggests that the resonances are associated with the excitation of the cross-polarized (quasi-TM) modes of the ridge, which in this case acts as a leaky rib waveguide~\cite{rib1,rib2}.

To verify this assumption, let us estimate the effective refractive indices of the quasi-TM modes of the ridge using the EIM [see Fig.~\ref{fig:1}(b); the used refractive indices are shown in parentheses]~\cite{pollock,pocket}.
Under this approach, the estimates of the wave numbers $k_\mathrm{y,TM,rib}$ can be found by solving the dispersion relation of the symmetric three-layer slab waveguide~\cite{pocket}:
\begin{equation}
\label{eq:disprel}
\tan(k w) = \frac{2 k \gamma}{k^2 - \gamma^2},
\end{equation}
where $k = \sqrt{k_0^2 n_\mathrm{TM,r}^2 - k_\mathrm{y,TM,rib}^2}$ and $\gamma = \sqrt{k_\mathrm{y,TM,rib}^2 - k_0^2 n_\mathrm{TM}^2}$.
The excitation condition of the found rib waveguide modes, which relates $w$ and $\theta$, reads as 
$k_\mathrm{y,TM,rib} = k_0 n_\mathrm{TE} \sin\theta$. 
The values $(w, \theta)$ satisfying this condition are shown with dashed red curves in Fig.~\ref{fig:2}(b), which are close to the location of the resonances.
Thus, the observed high-Q resonances are indeed associated with the excitation of the cross-polarized modes of the ridge.

\subsection{Bound states in the continuum}

\begin{figure}[ht!]
\centering\includegraphics{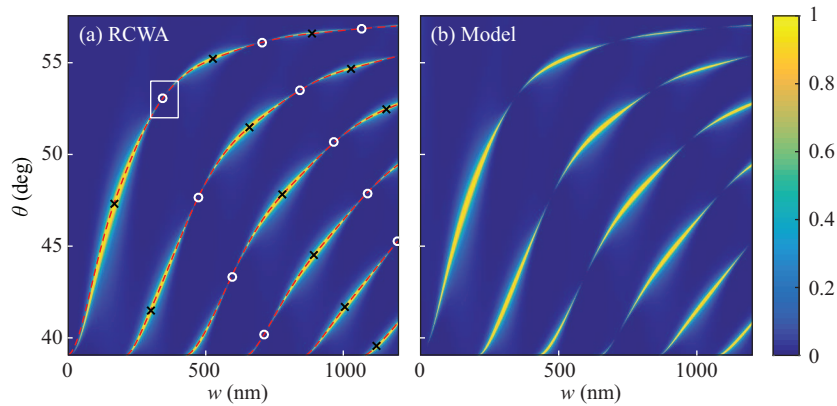}
\caption{\label{fig:5}(a)~The magnified fragment of the rigorously calculated reflectance spectrum shown in Fig.~\ref{fig:2}(a).
The red dashed curves show the mode dispersion obtained from the model of Section~\ref{sec:CWT} using Eq.~\eqref{dispeq}.
The $(w, \theta)$ points shown with the white circles and black crosses were obtained from Eqs.~\eqref{eq:bic:thetah} and~\eqref{eq:bic1:phases}.
The white circles predict the BIC locations, whereas the black crosses correspond to the ``low-Q'' resonances.
(b)~The model reflectance spectrum $|R|^2$ calculated using Eq.~\eqref{eq:r}.
}
\end{figure}

Figure~\ref{fig:5}(a) shows a magnified fragment of Fig.~\ref{fig:2}(a) corresponding to the angular range $\theta_\mathrm{TM} < \theta < \theta_\mathrm{TM,r}$. It is evident that the angular width (and, consequently, the quality factor) of the resonances strongly varies along the dispersion curves,
apparently vanishing at certain $(w,\theta)$ values marked with white circles in Fig.~\ref{fig:5}(a).
This phenomenon indicates the potential presence of the bound states in the continuum (BICs) in the considered structure.
Let us remind that the BICs are the eigenmodes of the structures with open diffraction channels, which, however, have an infinite quality factor and a real frequency due to symmetry reasons or parameter tuning.

\begin{figure}[ht!]
\centering\includegraphics{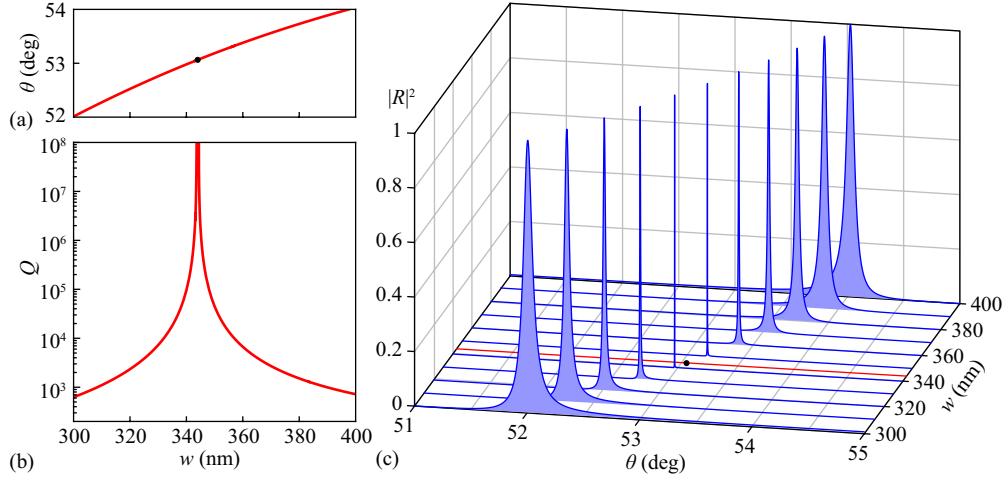}
\caption{\label{fig:Q}Rigorously calculated excitation angle~$\theta$~(a) and the quality factor~$Q$~(b) of the mode vs. the ridge width~$w$.
Angular reflectance spectra of the structure at different ridge widths~(c).
The red line shows the spectrum of the structure supporting a BIC.
The black points in~(b) and~(c) depict the BIC location.
}
\end{figure}

In order to determine whether the structure actually supports BICs, let us investigate the quality factor of the modes in the region shown by the white rectangle in Fig.~\ref{fig:5}(a).
In Fig.~\ref{fig:Q}, we present the resonance angle $\theta$ and the quality factor~$Q$ of the eigenmodes of the ridge,
which were calculated using the rigorous RCWA-based approach proposed by some of the present authors in~\cite{my:bykov:2013:jlt, my:bykov:2015:co} (see the details in Appendix~A). It is evident from Fig.~\ref{fig:Q} that at $w = 344$~nm, the quality factor diverges. This indicates the presence of a BIC. 

Figure~\ref{fig:Q}(c) shows the angular reflectance spectra of the structure at different ridge widths. 
According to Fig.~\ref{fig:Q}, the quality factor of the resonance increases when approaching the BIC, which results in a narrowing of the resonant reflectance peak.
In addition to the existence of a BIC, the spectra in Fig.~\ref{fig:Q}(c) also demonstrate that the investigated structure can be used as a narrow band spatial (angular) filter.
The FWHM of the reflectance peak in Fig.~\ref{fig:Q} varies from $0^\circ$ (at the BIC) to $0.1^\circ$.
Further calculations not presented here show that the same structure can also be used as a spectral filter.

In contrast to the positions of the resonances, the variation of the quality factor and the appearance of the BICs cannot be explained by an EIM-based approach.
Let us note, that the existence of BICs in a similar structure was predicted in~\cite{Zou}, where the ridge was considered as a photonic rib waveguide with an unconventional (BIC-based) guiding mechanism.
In contrast to~\cite{Zou}, we consider this structure as an integrated optical element for slab waveguide modes and, in the next section, derive a simple and accurate coupled-wave model describing its reflectance and transmittance spectra, predicting the high-Q resonances in the ridge, and confirming the existence of the BICs.

\section{Coupled-wave analysis of the resonances}\label{sec:CWT}

In this section, we present a simple model based on the coupled-wave equations, which explains the formation of high-Q resonances and predicts the location of the bound states in the continuum supported by the considered structure.

\subsection{Coupled-wave equations}

\begin{figure}[ht!]
\centering\includegraphics{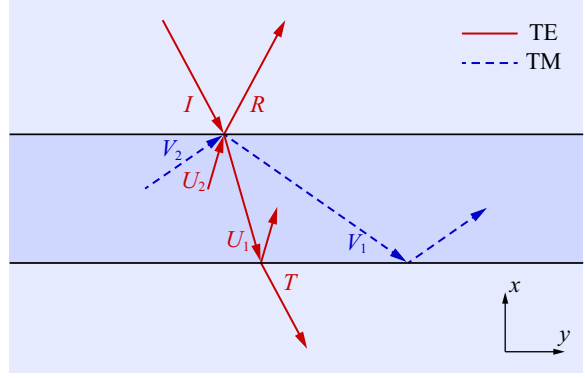}
\label{fig:CW}
\caption{\label{fig:cw}Plane wave diffraction by a slab supporting two cross-polarized waves coupled at the interfaces.}
\end{figure}

In order to obtain the coupled-wave equations, let us revisit the effective index model of Fig.~\ref{fig:1}(b), where 
the integrated structure shown in Fig.~\ref{fig:1}(a) is replaced with a $z$-invariant slab.
In contrast to this model considered above in Subsection~\ref{ssec:tandr}, here we will take into account the cross-polarization mode coupling which takes place in the considered integrated structure.
To do this, we represent the field inside the slab as TE and TM plane waves, which are coupled at the interfaces.
The field over and under the slab contains the incident, reflected, and transmitted TE waves only (see Fig.~\ref{fig:cw}).
The wave numbers of the plane waves are defined by the effective refractive indices: $n_\mathrm{TE,r}$ and $n_\mathrm{TM,r}$ for the waves inside the slab; $n_\mathrm{TE}$ for the waves outside the slab.

According to Fig.~\ref{fig:cw}, let us introduce the following plane waves:
the incident wave (with the complex amplitude $I$);
the reflected wave ($R$);
the TE and TM waves propagating downwards inside the slab ($U_1$ and $V_1$);
and the upward-propagating TE and TM waves inside the slab ($U_2$ and $V_2$).
The amplitudes $U_1$ and $V_1$ are defined at the lower interface of the slab;
the amplitudes of the same waves at the upper interface equal $U_1 \ee^{-\ii \phi}$ and $V_1 \ee^{-\ii \psi}$.
Similarly, the amplitudes $U_2$ and $V_2$ are defined at the upper interface of the slab;
the amplitudes of these waves at the lower interface are $U_2 \ee^{-\ii \phi}$ and $V_2 \ee^{-\ii \psi}$.
The phases $\phi$ and $\psi$ are defined through the effective refractive indices of the TE and TM waves inside the slab, respectively:
\begin{equation}
\label{eq:phi}
\begin{aligned}
\phi &= w \sqrt{k_0^2 n_\mathrm{TE,r}^2 - k_y^2},\\
\psi &= w \sqrt{k_0^2 n_\mathrm{TM,r}^2 - k_y^2}.\\
\end{aligned}
\end{equation}
Here, $w$ is the slab thickness, which is equal to the width of the ridge in the initial ``integrated'' diffraction problem; 
and $k_y$ is the wave vector component of the plane waves parallel to the slab interfaces.
Note that the square roots in Eq.~\eqref{eq:phi} are positive numbers that are the $x$-components of the wave vectors of the TE and TM plane waves inside the slab.

At the upper interface, the incident and scattered plane waves are related by a $3\times3$ scattering matrix:
\begin{equation}
\label{eq:smatrix}
\bmat{
U_1 \ee^{-\ii \phi}\\
V_1 \ee^{-\ii \psi}\\
R
}
=
\bmat{
r_1 & r_c & t\\
r_c & r_2 & t_c\\
t & t_c & r
}
\cdot
\bmat{
U_2\\
V_2\\
I
},
\end{equation}
where $r$ and $t$ are the reflection and transmission coefficients of the incident TE plane wave $I$;
$r_1$ and $r_2$ give the reflection coefficients of the $U$ and $V$ waves;
$r_c$ and $t_c$ are the cross-polarization reflection and transmission coefficients.
Note that the scattering matrix in Eq.~\eqref{eq:smatrix} is symmetric due to reciprocity.
Besides,  the scattering matrix is unitary due to energy conservation, since no absorption and no out-of-plane scattering occur in the considered integrated structure.
Let us note that the elements of the scattering matrix can be rigorously calculated using the RCWA~\cite{rcwa1,rcwa2,rcwa3,pml} by solving the 
problem of diffraction of a guided mode on an interface between two slab waveguides with different thicknesses: $h_\mathrm{wg}$ and $h_\mathrm{wg}+h_\mathrm{r}$.

By applying the same scattering matrix at the lower slab interface, we obtain the following set of coupled-wave equations:
\begin{equation}
\label{eq:cwe}
\left\{
\begin{aligned}
U_1 &= \ee^{\ii \phi} \lt( r_1 U_2 + r_c V_2 + t I \rt), \\
V_1 &= \ee^{\ii \psi} \lt( r_c U_2 + r_2 V_2 + t_c I\rt), \\
U_2 &= \ee^{\ii \phi} \lt( r_1 U_1 + r_c V_1 \rt), \\
V_2 &= \ee^{\ii \psi} \lt( r_c U_1 + r_2 V_1 \rt), \\
R &= t U_2 + t_c V_2 + r I.
\end{aligned}
\right.
\end{equation}
For simplicity, we omitted the equation for the transmitted wave amplitude $T$. However, the following theory can be derived for the analysis of the transmission coefficient as well.

By eliminating $U_1$, $U_2$, $V_1$, and $V_2$ from the system~\eqref{eq:cwe}, we represent the reflection coefficient $R$ in the following form:
\begin{equation}
\label{eq:r}
R = r+ \frac{\mathcal{N}}{\mathcal{D}_1 \mathcal{D}_2}.
\end{equation}
Here, the denominator is the product of the following two terms:
\begin{equation}
\label{eq:d1}
\mathcal{D}_1 = \lt(\ee^{-\ii \phi} + r_1\rt)\lt(\ee^{-\ii \psi} + r_2\rt) - r_c^2,
\end{equation}
\begin{equation}
\label{eq:d2}
\mathcal{D}_2 = \lt(\ee^{-\ii \phi} - r_1\rt)\lt(\ee^{-\ii \psi} - r_2\rt) - r_c^2,
\end{equation}
whereas the numerator $\mathcal{N}$ reads as
\begin{equation}
\label{eq:n}
\begin{aligned}
\mathcal{N} = &r_1 t^2 \ee^{-2\ii \psi} + 2 r_c t_c t  \ee^{-\ii \phi} \ee^{-\ii \psi} + r_2 t_c^2 \ee^{-2 \ii \phi}
-\lt(r_1 r_2 - r_c^2 \rt) \lt(r_2 t^2 - 2 r_c t_c t + r_1 t_c^2\rt).
\end{aligned}
\end{equation}
Here, we assume that the incident wave has unity amplitude ($I = 1$).

The modes of the structure correspond to the poles of the reflection coefficient~\cite{gippius, my:bykov:2013:jlt}.
Therefore, we can obtain the dispersion equation of the modes by equating the denominator in Eq.~\eqref{eq:r} to zero:
\begin{equation}
\label{dispeq}
\mathcal{D}_1 \mathcal{D}_2 = 0.
\end{equation}
The fact that the denominator is a product of two terms means that the considered structure supports two uncoupled sets of modes, namely, the symmetric and antisymmetric modes of the structure. 
One can calculate these modes independently by solving the equations ${\mathcal D}_1 = 0$ and ${\mathcal D}_2 = 0$. 

Let us consider the antisymmetric modes defined by equation ${\mathcal D}_1 = 0$.
If no cross-polarization coupling occurs (if $r_c = 0$), this equation has two independent solutions, 
$\ee^{-\ii \phi} + r_1 = 0$ and $\ee^{-\ii \psi} + r_2 = 0$, which correspond to the antisymmetric TE and the antisymmetric TM modes, respectively.
If $r_c \neq 0$, these modes are coupled and form antisymmetric modes with a more complex polarization state.
Similarly, the equation ${\mathcal D}_2 = 0$ describes the coupling of symmetric TE modes with symmetric TM modes.

It is noteworthy that the equation $\mathcal{D}_1 \mathcal{D}_2 = 0$ can be written in the following matrix form:
$$
\det \lt\{
\mat{I} - 
\lt(\bmat{r_1 & r_c\\
r_c & r_2}
\cdot
\bmat{\ee^{\ii \phi} &0 \\
0&\ee^{\ii \psi}}
\rt)^2\rt\}
= 0.
$$
This expression is used as the dispersion equation in the theory of the so-called matrix Fano resonances of high-contrast gratings~\cite{matrixfp}.
A similar study of high-contrast gratings was presented in~\cite{orta2}, where the authors used the concept of generalized Fabry--P{\'e}rot interferometers~\cite{orta1}.

\subsection{Bound states in the continuum}
Here, we show that the coupled-wave model derived in the previous subsection predicts the bound states in the continuum (BICs) in the considered structure.
In what follows, we derive explicit expressions for the frequency $\omega$ and wave number $k_y$ of the BIC, as well as the expressions for the ridge width $w$ and the angle of incidence $\theta$ describing the BIC locations in Fig.~\ref{fig:5}(a).

The modes of the structure are described by the zeros of the denominator in Eq.~\eqref{eq:r}.
Therefore, the BICs are real-$\omega$, real-$k_y$ solutions of one of the following equations: $\mathcal{D}_1 = 0$ or $\mathcal{D}_2 = 0$.
Despite the simple form of these equations, it is difficult to solve them for real $\omega$ and $k_y$. Instead of doing this, we will use an alternative approach.

Assume that $\omega$ is the real frequency of a certain BIC.
Since the modes are described by the zeros of the denominator in Eq.~\eqref{eq:r}, the reflection coefficient~\eqref{eq:r} at frequency $\omega$ tends to infinity.
This apparently violates the energy conservation condition. 
To overcome this contradiction, $\omega$ should also be a zero of the numerator $\mathcal{N}$ in Eq.~\eqref{eq:r}.
In this case, the pole of $R$, which is associated with the zero of the denominator, will be compensated by the zero of the numerator.
We will use this fact to find the BICs.

Let us first consider the modes defined by the equation $\mathcal{D}_1 = 0$.
To find the corresponding BICs, we solve the following system with respect to the exponents $\ee^{\ii \phi}$ and $\ee^{\ii \psi}$:
\begin{equation}
\label{eq:bic1}
\lt\{
\begin{aligned}
\mathcal{D}_1 &= 0,\\
\mathcal{N}&=0.
\end{aligned}
\rt.
\end{equation}
After some simple transformations, we find two different solutions of the system~\eqref{eq:bic1}.
The first solution reads as
\begin{equation}
\label{eq:bic1:good}
\ee^{\ii \phi} = \frac{t_c}{r_c t - r_1 t_c},\;\;\;
\ee^{\ii \psi} = \frac{t}{r_c t_c - r_2 t}.
\end{equation}
The second solution can be written as
\begin{equation}
\label{eq:bic1:bad}
\ee^{\ii \phi} = \pm\sqrt{\frac{r_2}{r_1 (r_1 r_2 - r_c^2)}},\;\;\;
\ee^{\ii \psi} = \mp\sqrt{\frac{r_1}{r_2 (r_1 r_2 - r_c^2)}}.
\end{equation}

Since both $\omega$ and $k_x$ have to be real for a BIC to take place, the phases~\eqref{eq:phi} should also be real; 
therefore, $\ee^{\ii \phi}$ and $\ee^{\ii \psi}$ should be unity-amplitude complex numbers.
To verify whether this is true, let us recall that $r$, $r_1$, $r_2$, $r_c$, $t$, and $t_c$ are not arbitrary complex numbers, but adhere to the energy conservation law, which implies the unitarity of the scattering matrix in Eq.~\eqref{eq:smatrix}.
Simple deductions based on the unitarity property suggest that the right-hand sides in Eqs.~\eqref{eq:bic1:good} have unity amplitudes (see Appendix~B), hence, Eq.~\eqref{eq:bic1:good} indeed describes the BICs.
This, however, is not the case for Eq.~\eqref{eq:bic1:bad}: by considering an arbitrary $3\times3$ unitary matrix, one can show that the absolute values of the right-hand sides of Eqs.~\eqref{eq:bic1:bad} are, as a rule, not equal to one.

Now, let us obtain explicit expressions for the phases $\phi$ and $\psi$.
To do this, we equate the arguments of the left- and right-hand sides of Eqs.~\eqref{eq:bic1:good}:
\begin{equation}
\label{eq:bic1:phases}
\phi = \pi m + \arg \frac{t_c}{r_c t - r_1 t_c},\;\;\;
\psi = \pi l + \arg \frac{t}{r_c t_c - r_2 t}.
\end{equation}
Here, both $m$ and $l$ are \emph{even} integer numbers.
The obtained Eqs.~\eqref{eq:bic1:phases} describe the positions of the BICs satisfying equation $\mathcal{D}_1 = 0$.
Similar analysis for the equation $\mathcal{D}_2 = 0$ results in the very same Eq.~\eqref{eq:bic1:phases} where both $m$ and $l$ are \emph{odd} integers.
Besides, according to Eq.~\eqref{eq:phi}, the phases $\phi$ and $\psi$ should be positive.
By defining the range of the principal value of the argument of a complex number as the $[0, 2\pi)$ interval, we impose the nonnegativity condition also on the $m$ and $l$ values.
Additionally, according to Eqs.~\eqref{eq:phi}, condition $\phi > \psi$ should be met since $n_\mathrm{TE,r} > n_\mathrm{TM,r}$.

Having obtained the expressions for $\phi$ and $\psi$, we can finally solve Eqs.~\eqref{eq:phi} for $\omega$ and $k_y$:
\begin{equation}
\label{eq:bic:omegakx}
\omega = \frac{\rm c}{w}\sqrt{\frac{\phi^2 - \psi^2}{n_\mathrm{TE,r}^2 - n_\mathrm{TM,r}^2}},\;\;\;
k_y = \frac{1}{w}\sqrt{\frac{\phi^2 n_\mathrm{TM,r}^2 - \psi^2 n_\mathrm{TE,r}^2}{n_\mathrm{TE,r}^2 - n_\mathrm{TM,r}^2}}.
\end{equation}
The obtained Eqs.~\eqref{eq:bic1:phases} and~\eqref{eq:bic:omegakx} describe the positions (frequencies and wave numbers) of the BICs for a given structure geometry.
If the frequency of the incident light is fixed, a BIC can be obtained by tuning the angle of incidence $\theta$ and the ridge width $w$, as it is evident from Fig.~\ref{fig:5}.
In this case, we recall the expression $k_y = k_0 n_\mathrm{TE} \sin\theta$ and solve Eqs.~\eqref{eq:phi} for $\theta$ and $w$:
\begin{equation}
\label{eq:bic:thetah}
w = \frac{\rm 1}{k_0}\sqrt{\frac{\phi^2 - \psi^2}{n_\mathrm{TE,r}^2 - n_\mathrm{TM,r}^2}},\;\;\;
\theta = \arcsin\lt(
\frac{1}{n_\mathrm{TE}}\sqrt{\frac{\phi^2 n_\mathrm{TM,r}^2 - \psi^2 n_\mathrm{TE,r}^2}{\phi^2 - \psi^2}}
\rt)
.
\end{equation}

It is important to note that the very existence of the analytical expressions~\eqref{eq:bic:omegakx} and~\eqref{eq:bic:thetah} means that the BICs in the considered structure are robust.
Indeed, any variation of the incident light wavelength and/or of the geometrical parameters of the structure results in a change in the right-hand sides of Eqs.~\eqref{eq:smatrix},~\eqref{eq:bic1:phases}, and~\eqref{eq:bic:thetah}.
Therefore, the BICs will not disappear but change their locations in the $(w, \theta)$ plane according to Eq.~\eqref{eq:bic:thetah}.
This reasoning is valid provided that only two scattering channels remain open (see Section~\ref{sec:2}).

Let us also note that since the BICs are simply real-$\omega$ solutions of the equation $\mathcal{D}_1 = 0$ (or $\mathcal{D}_2 = 0$), the BIC conditions are uniquely defined by the reflection coefficients $r_1$, $r_2$, and $r_c$ used in Eqs.~\eqref{eq:d1} and~\eqref{eq:d2}.
However, to obtain simple closed-form expressions for the BIC condition, we had to use the values of $t$ and $t_c$ in Eq.~\eqref{eq:bic1:phases}.

\subsection{Comparison with the rigorous simulations}
To verify the presented theoretical model, we used it to calculate the reflectance spectrum, the mode dispersion, and the locations of the BICs.
The obtained results were compared with the RCWA simulation results.

Figure~\ref{fig:5}(b) shows the reflectance spectrum calculated using Eqs.~\eqref{eq:r}--\eqref{eq:n}.
This model spectrum is in excellent agreement with the rigorously calculated one shown in Fig.~\ref{fig:5}(a).
Slight discrepancies between the two spectra can be seen only in the vicinity of $w=0$.
These discrepancies are due to the near-field interactions in a narrow ridge, which are neglected in the presented theoretical model.

The dispersion of the modes was calculated by numerically solving the equation $\mathcal{D}_1 \mathcal{D}_2 = 0$, where $\mathcal{D}_1$ and $\mathcal{D}_2$ are given by Eqs.~\eqref{eq:d1} and~\eqref{eq:d2}.
For the matter of illustration, we show the dispersion not in Fig.~\ref{fig:5}(b) but in the rigorously calculated Fig.~\ref{fig:5}(a).
The presented analytically calculated dispersion curves are in perfect agreement with the resonances in the rigorously calculated spectrum.
Using the model, we also calculated the quality factors of the modes, however, they are visually identical to the rigorously calculated curve in Fig.~\ref{fig:Q} and thus are not shown.

The white circles depicting the BICs in Fig.~\ref{fig:5}(a) were calculated using Eqs.~\eqref{eq:bic:thetah} and~\eqref{eq:bic1:phases}.
It is evident from the figure that the presented coupled-wave model provides an accurate estimate of the BIC positions.
As mentioned in the previous subsection, these white circles were obtained when $m$ and $l$ in Eq.~\eqref{eq:bic1:phases} have the same parity.
It is interesting to note that when $m$ and $l$ are of different parity, 
the corresponding $(w, \theta)$ points obtained from Eq.~\eqref{eq:bic:thetah} lie on the ``resonant'' curves $|R| = 1$;
this can be shown by direct substitution into Eq.~\eqref{eq:r}.
These point are shown with black crosses in Fig.~\ref{fig:5}(a).
In contrast to the BICs (infinite-Q resonances shown with white circles), the black crosses describe the relatively low-Q resonances of the structure.

In Section~\ref{sec:3}, using the EIM approach, we argued that the BICs appear on the quasi-TM dispersion curves of the ridge.
The derived theoretical model provides us with a deeper insight into the BIC formation mechanism.
According to the model, the BICs in the considered structure emerge due to the interaction of the quasi-TE and quasi-TM modes.
If we exclude the TE waves propagating inside the ridge from the presented coupled-wave model, the BICs will disappear from the model spectrum.

\section{Conclusion}\label{sec:conclusion}
In the present work, we demonstrated that in the case of diffraction of a TE-polarized mode of a slab waveguide by a dielectric ridge located on the waveguide surface, pronounced resonances in the reflectance and the transmittance spectra occur.
These resonances take place at oblique incidence of the mode when only two scattering channels are open, which correspond to the reflected and the transmitted mode of the same polarization.
The considered structure supports resonances of arbitrarily high quality factor, as well as modes with an infinite Q-factor, i.e.\ bound states in the continuum (BICs).

We developed a coupled-wave model explaining the optical properties of the structure.
By considering the interaction of the TE and TM modes inside the ridge, we derived an accurate expression for the complex reflection coefficient of the structure. 
Using the proposed model, we analytically confirmed that the considered structure indeed supports robust BICs, 
which can be obtained by tuning two parameters: the ridge width and the angle of incidence.
We also obtained simple closed-form expressions for these parameters providing the BICs.
The developed model is in perfect agreement with the presented full wave simulation results.

The authors believe that the investigated planar structure can be used as an integrated optical spatial (angular) or spectral filter.
The existence of high-Q resonances and BICs in the structure also makes it promising for sensing, enhancing non-linear light--matter interactions, and analog optical signal processing.
In particular, since the transmission coefficient of the ridge strictly vanishes at the resonances, the structure enables performing spatial optical differentiation in transmission as well as spatial optical integration in reflection.
This application of the structure will be the subject of a separate work.

\appendix
\section*{Appendix A. Details of the eigenmode calculation}\label{app:A}
In this appendix, we present the details of the numerical calculation of the eigenmodes of the structure.
In particular, we are interested  in the quality factor of the modes, which is defined as $Q = \Re \omega / (-2\Im\omega)$, where $\omega$ is the complex eigenfrequency.

The complex eigenfrequency $\omega$ can be found as the pole of the scattering matrix $\mat{S}(\omega, k_y)$~\cite{my:bykov:2013:jlt},
where $k_y$ is the wave number of the mode.
However, at a fixed angle of incidence $\theta$ the wave number itself depends on the angular frequency as $k_y = (\omega / {\rm c}) n_{\rm TE} \sin\theta$.
Here, $n_{\rm TE}$ is the effective refractive index of the incident mode, which also depends on the frequency $\omega$ due to material and structural dispersion.
Therefore, to calculate the complex frequency of the mode, we numerically find the complex pole of the matrix $\mat{S}\big(\omega, (\omega / {\rm c}) \cdot n_{\rm TE}(\omega) \sin\theta \big)$, which is considered as a function of $\omega$.
Note that the calculated mode will have both a complex frequency and a complex wave number. 
Therefore, it will decay both in time and in space.
The presented approach guarantees that the imaginary part of the complex frequency is accurately calculated and describes the line width of the resonance in the frequency spectrum.

It is worth noting that the calculation of the scattering matrix requires defining the basis of incident and scattered waves.
This requires special attention in the case of complex frequencies as it was discussed in~\cite{my:bykov:2015:co}.

Having a method for calculating the complex frequencies, and, hence, the complex wavelengths of the modes, we can numerically solve the equation $\Re\lambda = 630$~nm for real $\theta$, which gives us the method for calculating the angle of incidence $\theta$, at which the mode is excited exactly at the given wavelength.
Moreover, since we calculate the complex $\omega$ as the pole of the scattering matrix, we can also calculate the quality factor.
This is the approach we used to obtain Fig.~\ref{fig:Q}.

\section*{Appendix B. Unity amplitudes of the right-hand sides of Eqs.~\eqref{eq:bic1:good}}\label{app:B}
Here, we show that the right-hand sides of Eqs.~\eqref{eq:bic1:good} have unity amplitude.
To do this, we consider the scattering matrix $\mat{S}$ used in Eq.~\eqref{eq:smatrix}.
We write the inverse of $\mat{S}$ in two ways: using the unitarity property ($\mat{S}^{-1} = \mat{S}^{*}$) and in terms of the adjugate matrix.
By equating these two matrices, we obtain
$$
\bmat{
r_1^* & r_c^* & t^*\\
r_c^* & r_2^* & t_c^*\\
t^* & t_c^* & r^*
}
=
\frac{1}{\det\mat{S}}
\bmat{
 r r_2 - t_c^2 & t t_c-r r_c & r_c t_c-r_2 t\\
 t t_c - r r_c& r r_1 - t^2& r_c t - r_1 t_c\\
 r_c t_c -r_2 t& r_c t - r_1 t_c& r_1 r_2 - r_c^2}.
$$
Now we can equate the absolute values of the corresponding matrix elements in this equation.
Since the determinant of a unitary matrix lies on the unit circle, we arrive at the equalities $|t_c| = |r_c t - r_1 t_c|$ and $|t| = | r_c t_c-r_2 t|$, which prove the required statement.

\section*{Funding}
Russian Foundation for Basic Research (16-29-11683, 17-47-630323); Ministry of Science and Higher Education of the Russian Federation (007-GZ/Ch3363/26).

\section*{Acknowledgments}
Numerical investigation of the optical properties of the structure was supported under Russian Foundation for Basic Research project 16-29-11683, the analysis of the resonances based on the effective index method was supported under Russian Foundation for Basic Research project 17-47-630323, and the derivation of the coupled-wave model and the analysis of the BIC formation mechanism was supported by Ministry of Science and Higher Education of the Russian Federation.


\end{document}